\documentclass[runningheads]{llncs}
\usepackage{graphicx}
\usepackage{subfigure}
\usepackage{amsmath}
\usepackage{multirow}
\usepackage{hyperref}
\hypersetup{
    colorlinks=true,
    linkcolor=blue,
    filecolor=blue,      
    urlcolor=blue,
    citecolor=cyan,
}
\begin{document}
\title{Renal Cell Carcinoma Detection and Subtyping with Minimal Point-Based Annotation in Whole-Slide Images}
%
%

\author{Zeyu Gao\inst{1,2} \and
Pargorn Puttapirat\inst{1,2} \and
Jiangbo Shi\inst{1,2} \and
Chen Li\inst{1,2}}

\authorrunning{Z. Gao et al.}

%
\institute{National Engineering Lab for Big Data Analytics, Xi'an Jiaotong University, Xi'an, Shaanxi 710049, China \and
School of Electronic and Information Engineering, Xi'an Jiaotong University, Xi'an, Shaanxi 710049, China \\
\email{gzy4119105156@stu.xjtu.edu.cn}}

\titlerunning{RCC Detection and Subtyping with Min-Point Annotation in WSIs}
\maketitle              
\begin{abstract}

Cancerous region detection and subtyping in whole-slide images (WSIs) are fundamental to renal cell carcinoma (RCC) diagnosis. The main challenge in the development of automated RCC diagnostic systems is the lack of large-scale datasets with precise annotations. In this paper, we propose a framework that employs a semi-supervised learning (SSL) method to accurately detect cancerous regions with a novel annotation method called Minimal Point-Based (Min-Point) annotation. The predicted results are efficiently utilized by a hybrid loss training strategy in a classification model for subtyping. The annotator only needs to mark a few cancerous and non-cancerous points in each WSI. Experiments on three significant subtypes of RCC proved that the performance of the cancerous region detector trained with the Min-Point annotated dataset is comparable to classifiers trained on the dataset with full cancerous region delineation. In subtyping, the proposed model outperforms a model trained with only whole-slide diagnostic labels by 12\% in terms of the testing f1-score. We believed that our “detect then classify” schema combined with the Min-Point annotation would set a standard for developing intelligent systems with similar challenges, especially in cancer research.

\end{abstract}
\section{Introduction}
Renal cell carcinoma (RCC) accounts for more than 90\% in the kidney cancer cases. While it contains roughly ten histologic and molecular subtypes \cite{hsieh2017renal}, the three including Clear cell (cc), papillary (p), and chromophobe (ch) are major \cite{ljungberg2015eau}. RCC diagnosis is vital and generally consists of a histologic subtype and a grade. Each subtype could yield dramatically different prognoses, treatment strategies, and survival outcomes \cite{soyer1997renal,delahunt1997papillary,megumi1998chromophobe}. Similarly, the five-year overall survival outcome varies from 32\% to 91\% for ccRCC with different grades \cite{ljungberg2015eau}. Tumor grading, either with the Fuhrman's \cite{fuhrman1982prognostic} or the ISUP grading system \cite{isup}, are developed based on the visual inspection of cancerous cells hence identifying cancerous regions is critical \cite{golatkar2018classification,kong2018invasive,le2019pancreatic}. These situations have led us to focus on the development of accurate cancerous region detection and subtyping in RCC.

Deep learning is a method of choice for such problem. It requires an extraordinary amount of manually labeled data as seen in the Camelyon 2016 \cite{Camelyon} and TUPAC 2016 \cite{TUPAC16} datasets. Nevertheless, keep annotating large-scale datasets for other types of cancer is very expensive. Existing works rely on complete and precise annotations to develop cancer region detection. To reduce annotation efforts, we propose a novel annotation method called Min-Point annotation that only requires annotators to minimally annotate points on cancerous and non-cancerous regions of WSI. 

Regarding the Min-Point annotation, we adopt a semi-supervised learning (SSL) strategy for cancer region detection. SSL has been proved to be effective in utilizing unlabeled data, thus reducing the need for large-scale annotation in natural images \cite{ssl_m1,ssl_m2,ssl_m3} and medical images \cite{ssl_mm1,ssl_mm2,ssl_mm3}. The proposed detection framework utilizes a holistic SSL approach called MixMatch. Berthelot D et al. \cite{berthelot2019mixmatch} has unified current SSL approaches to form a new algorithm which obtains state-of-the-art results on several natural image datasets. Our work takes WSIs with Min-Point annotation as the training data to build the initial classifier. Then, a relatively larger set of unlabeled WSIs called ‘extension set’ was used to fine-tune the model, resulting in the final classifier. After that, based on the detected cancerous regions, we proposed a hybrid loss to train a robust deep learning subtype classifier. The proposed approach is different from \cite{tabibu2019pan} since the latter only trains the classifier with diagnosis provided by The Cancer Genome Atlas (TCGA), ignoring negative effects of non-cancerous regions in WSIs and being unable to give evidential outputs.

In this paper, our main contributions are as the following. {\bfseries(1)} We introduced a cancer region detection framework for RCC with a novel annotation method called Min-Point annotation, combining with the technique proposed in \cite{berthelot2019mixmatch}. The proposed model can be applied to other large-scale unlabeled dataset with minimal modifications. {\bfseries(2)} We proposed a subtyping framework for RCC based on detected cancerous regions by formulating a hybrid loss for the RCC subtype classifier. {\bfseries(3)} Our experiments on three major subtypes of RCC from TCGA prove the effectiveness of the proposed framework. The cancer region detection model was only trained with a relatively small set of the labeled data and a large set of unlabeled data when compared to those required by fully-supervised models trained on the manually annotated dataset. Our subtype classifier has 12\% better f1-score than the baseline. As an extra feature, the framework can provide evidential information along with its classification output.

\section{Proposed Framework}
\subsection{Minimal Point-Based Annotation}

\begin{figure}[t]
\centering
\subfigure[]{
\begin{minipage}[b]{0.45\textwidth}
\centering
\includegraphics[width=0.45\textwidth]{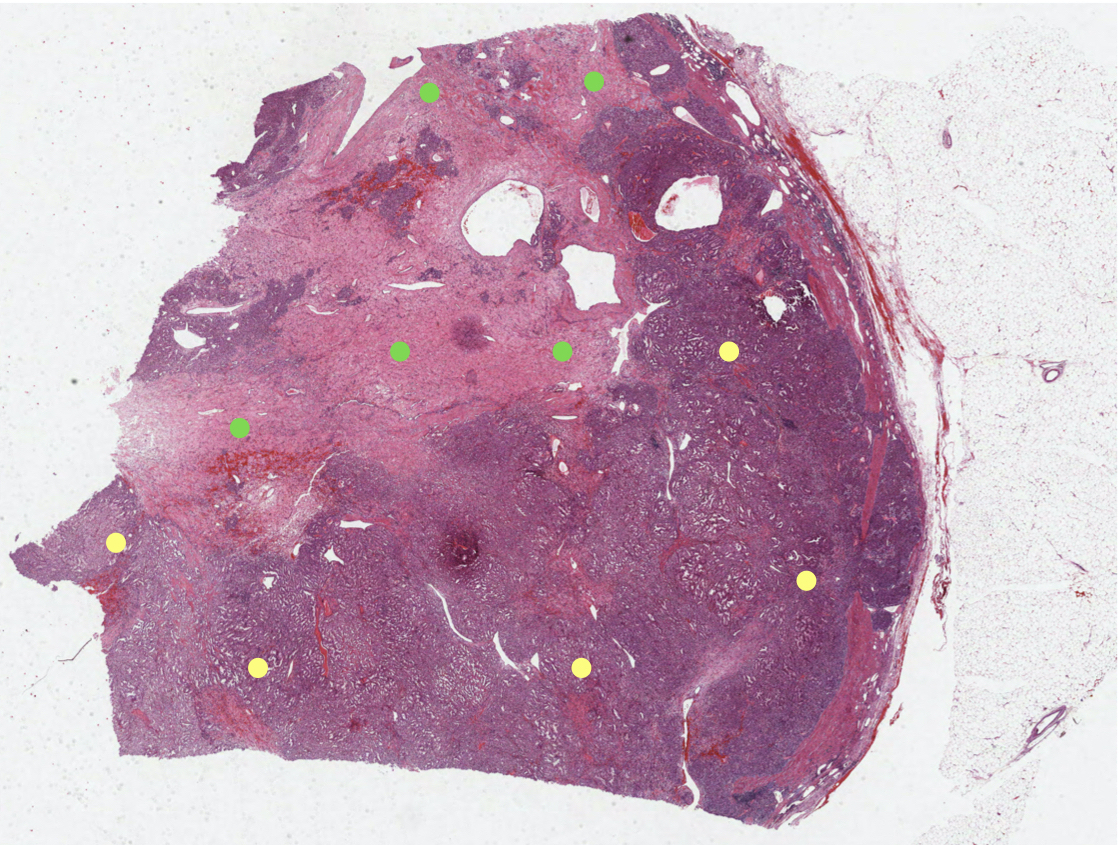}
\includegraphics[width=0.45\textwidth]{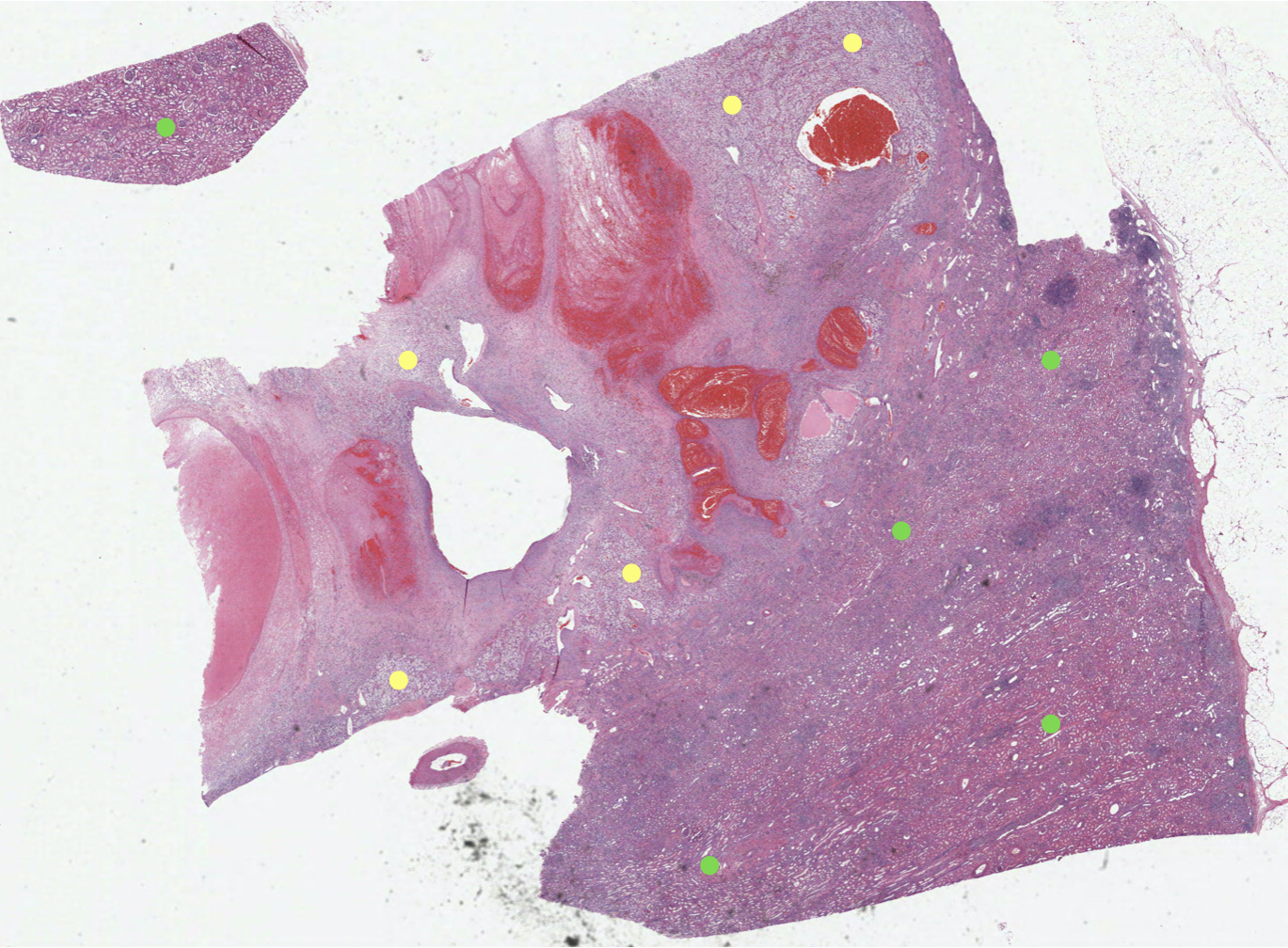}
\end{minipage}}
\subfigure[]{
\begin{minipage}[b]{0.45\textwidth}
\centering
\includegraphics[width=0.45\textwidth]{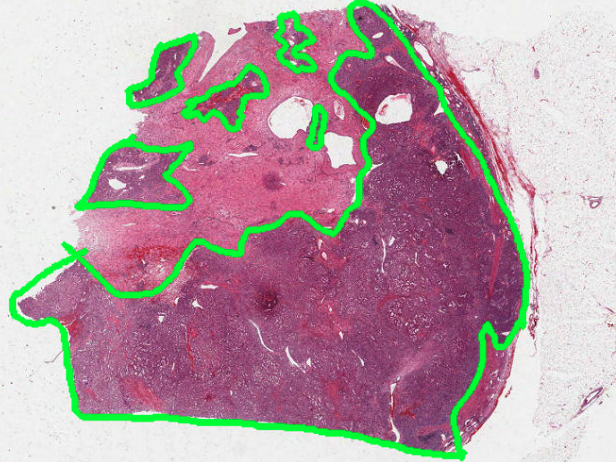}
\includegraphics[width=0.45\textwidth]{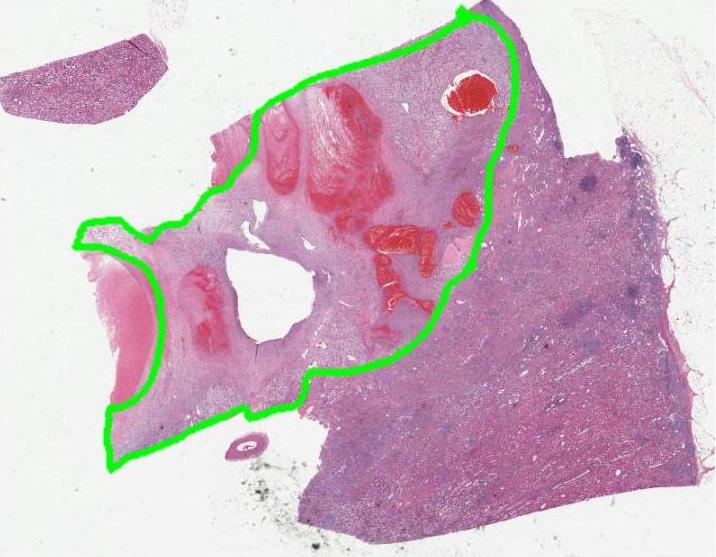}
\end{minipage}}
\caption{The Minimal Point-Based annotations (a) consist of positive points (yellow) and negative points (green) which indicate cancerous and non-cancerous regions. The complete region annotation (b) consist of one or multiple enclosed regions with precise perimeters on cancerous regions and non-cancerous region outside.} \label{fig1}
\end{figure}

Typically, the development of a cancer region detection model requires datasets where annotators must provide accurate boundaries of all cancer regions (complete region annotation) as shown in Fig. \ref{fig1}(b). To achieve precise perimeters, annotators need to zoom in and out between high and low magnification and outline the regions carefully. This type of activity is time consuming, and has to be done by domain experts. With Min-Point annotation, experts such as pathologists would only need to add a few points scattered on both cancerous and non-cancerous regions of WSI at low magnification by OpenHI \cite{puttapirat2019openhi-real} as shown in Fig. \ref{fig1}(a). The number of the points could be as few as five to ten. According to our annotation experience, Min-Point annotation can reduce the annotation time to roughly one-twentieth when compared to the complete annotation.

The minimally annotated dataset was used to generate labeled and unlabeled data. The labeled patches including cancerous and non-cancerous patches were created based on annotated points positioned at the center. The unlabeled patches were extracted by the sliding window strategy with a background filter. Following the experience of pathologists, the original size of a patch was $2000\times 2000$ pixels at 40x magnification, then, the patches were resized to $224\times 224$ pixels for training. Code is available at: \url{https://gitlab.com/BioAI/RCC_DS}.

\subsection{SSL model For Cancer Region Detection}

\begin{figure}[t]
\centering
\includegraphics[width=\textwidth]{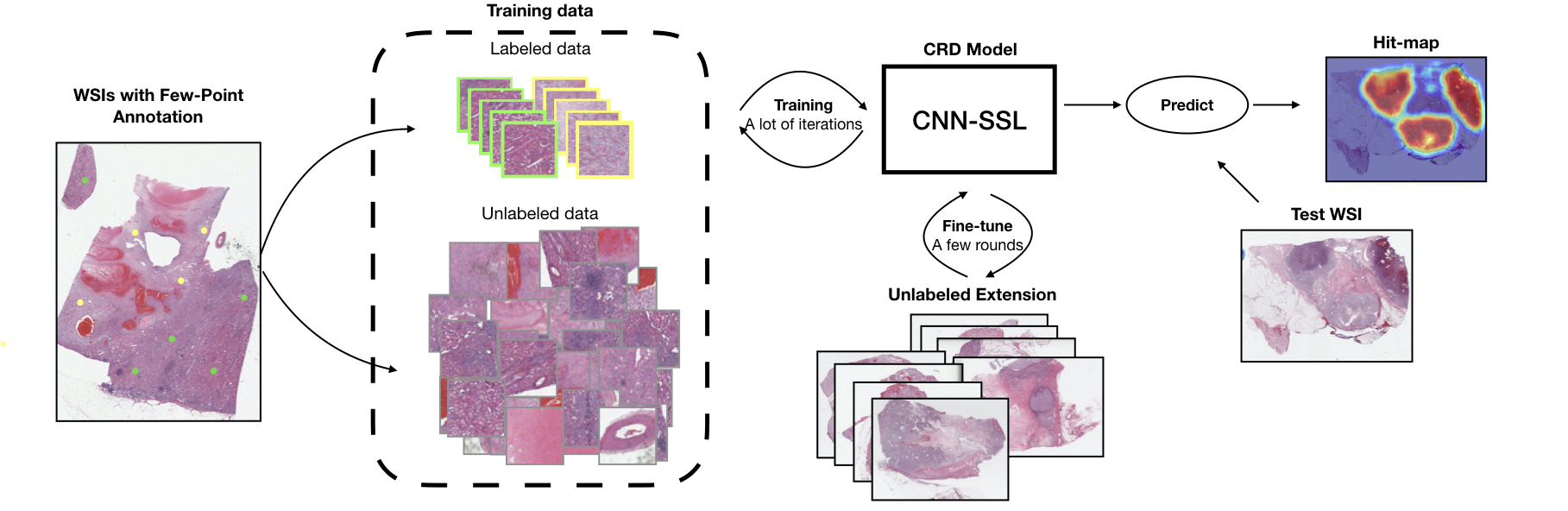}
\caption{The proposed semi-supervised cancer region detection framework} \label{fig2}
\end{figure}

With Min-Point annotation, insufficient training data is expected. This is where we would exploit valuable information from large amounts of unlabeled data. SSL becomes a natural choice. The SSL approach which combines multiple methods called MixMatch proposed in \cite{berthelot2019mixmatch} has proven to be effective in leveraging unlabeled data. We adopt this approach and integrate it into the proposed framework. First, Berthelot D et al. augment the labeled and unlabeled data, specifically $K$ augmentations is used on unlabeled data. Then, they use a classification model to get a "guess" label, which performed a sharpening \cite{goodfellow2016deep} to lower the entropy for each augmented unlabeled data. After that, MixUp \cite{zhang2017mixup} is utilized for both labeled and unlabeled samples. Finally, a combined loss for SSL is minimized for model training.

Let ${D_{l}}$ and ${D_{u}}$ be the set of labeled and unlabeled data, ${{D_{l}}}'$ and ${{D_{u}}}'$ be the set of augmented labeled and unlabeled data, respectively. $C$ stands for the numbers of classes, $H(p,q)$ is the cross-entropy between two distributions and $p_{pred}$, $q_{pred}$ represents the probability vectors of the model outputs. The combined loss function is as the following:

\begin{equation}
{{D_{l}}}',{{D_{u}}}'= MixMatch({D_{l}},{D_{u}})
\end{equation}
\begin{equation}
L=\frac{1}{\left | {D_{l}}' \right |}\sum_{x,p\in D_{l}}H\left ( p,p_{pred} \right )+\lambda \frac{1}{C\left | {D_{u}}' \right |}\sum_{u,q\in {D_{u}}'}\left \| q-q_{pred} \right \|_{2}^{2}
\end{equation}

\noindent Notice that $\lambda$ controls the importance of unlabeled data loss. In our experiment, we increase $\lambda$ linearly to its maximum over the training process. Since the model is unstable at the beginning, then becomes more discriminative as the iterations continue.  In the fine-tuning phrase, we want the model to learn more knowledge from the other part of unlabeled data (the extension set), so we set $\lambda$ as a constant. The data in the extension set is completely unlabeled, and it is hard for the model to learn valuable knowledge from this data. Still, if the model is already discriminative, the unlabeled data could help to increase its robustness. The overall cancer region detection framework, as shown in Fig.~\ref{fig2}, includes (1) labeled and unlabeled image patches generate from minimally annotated WSIs for training, (2) training of a convolutional neural network with SSL strategy (CNN-SSL) and (3) involving an extension dataset consist of unlabeled data to fine-tune the CNN-SSL model.

\subsection{Hybrid Loss for Subtyping}
\begin{figure}
\centering
\includegraphics[width=\textwidth]{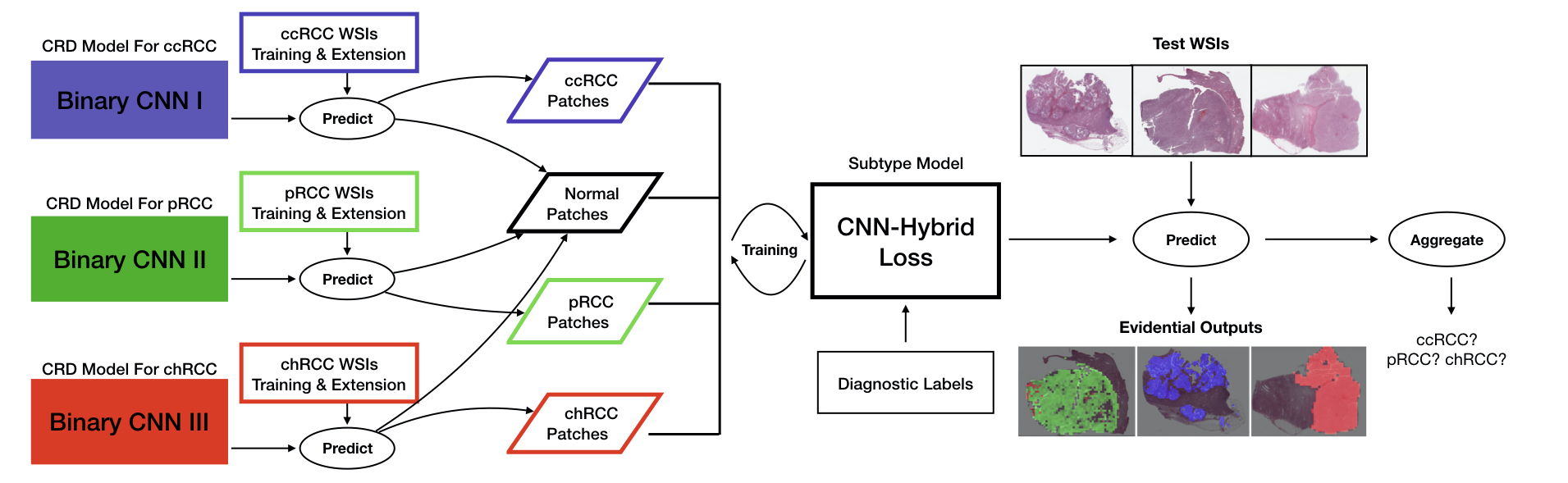}
\caption{Proposed subtyping framework} \label{fig3}
\end{figure}

After cancerous regions are recognized, they could be used for subtyping. Tabibu S et al. \cite{tabibu2019pan} treat the subtyping as a three-class classification problem (ccRCC vs. pRCC vs. chRCC) without considering the non-cancerous regions. Since there is not much difference between non-cancerous tissues among different subtypes, labeling these normal tissues as corresponding subtypes will have an adverse impact on both training and prediction of the model. We consider the non-cancerous regions as an additional "normal" class to form a four-class classification (normal vs. ccRCC vs. pRCC vs. chRCC), as shown in Fig.~\ref{fig3}. The binary CNN I, II, III are cancer region detection models for ccRCC, pRCC, chRCC, respectively. After predicting the labels of each patch, an evidential result can be presented to pathologists, and then the subtype label can be obtained by aggregation operation. In our experiment, we use majority voting to form the final decision without considering the predicted normal patches.

Regarding the discarded predicted normal patches, we strengthen the constraint on misclassified patches between subtypes, i.e. we increase the loss value of patches classified to other subtypes to achieve a better subtyping performance. For example, a patch from ccRCC WSI is classified to pRCC or chRCC. If this patch is classified as normal or ccRCC, it would be ignored. In doing so, the penalty for cross-subtypes false is increased by using the hybrid loss, which consists of a standard cross-entropy loss and a novel subtype loss. Each patch was assigned a subtype based on the slide-level diagnostics of the corresponding WSI.

There are two labels for each patch $x$, one is a four-class label $y$, the other is a three-class label $z$. Let $D$ be the set of training data, $p_{pred}=[p_{0},p_{1},...,p_{C}]$ stands for the output probabilities of $C$ classes, class 0 indicates normal class. We transform $p_{pred}$ to a subtype probability vector $s_{pred}$ as (\ref{subtype-pred}). Then, the hybrid loss function is defined as (\ref{hybrid-loss}) where $p,s$ represent one-hot codes of $y,z$, respectively. $\mu$ is a hyperparameter. As $\mu\rightarrow 0$, the loss function will return to traditional cross-entropy loss, raising $\mu$ encourages the model to increase the penalties for misclassification between different subtypes and reduce the sensitivity to some noise about "normal" and "cancer" class. In our experiment, we set $\mu=5$.
\begin{equation}
s_{pred}=\begin{cases}
 & \ \ \ s_{i}=p_{i}\ \ \ \   \text{ if } i\neq z \\ 
 & s_{i}=p_{i}+p_{0}\ \text{ if } i= z
\end{cases}
\label{subtype-pred}
\end{equation}
\begin{equation}
L=\frac{1}{\left | D \right |}\left ( \sum_{x,p,s\in D}H\left ( p,p_{pred} \right ) + \mu H(s,s_{pred}))\right )
\label{hybrid-loss}
\end{equation}

\section{Experiment}

\subsection{Datasets}

\begin{table}[t]
\caption{Dataset statistics}
\label{tab1}
\centering
\begin{tabular}{|c|c|c|c|c|c|}
\hline
WSIs(Patches)  & Training & Extension & Validation & Test & Total \\ \hline
ccRCC Detection & 20 (200,19999)       & 58 (59216)        & 10 (11201)        & 35 (35286)  & 123    \\ \hline
pRCC Detection  & 15 (150,13585)       & 48 (45953)        & 5 (4361)         & 20 (14953)   & 88    \\ \hline
chRCC Detection & 5 (50,6192)        & 25 (25058)        & 5 (5093)         & 10 (10593)   & 45    \\ \hline
Subtyping       & 171 (170003)      & -         & 20 (20655)         & 65 (60832),463  & 654   \\ \hline
\end{tabular}
\end{table}

A total of 654 WSIs (299, 254, 101 for three subtypes) of RCC from TCGA were used. All selected slides were scanned at 40x magnification. Three sets of data ($D_{ccRCC}$, $D_{pRCC}$, $D_{chRCC}$) were created for cancer region detection. Two pathologists were asked to annotate 10 points (5 cancerous, 5 non-cancerous) on each WSIs for training. Training sets consisted of a few labeled, a large number of unlabeled patches and a relatively larger unlabeled set (extension set) was used for fine-tuning. For subtyping, dataset $D_{subtype}$ consist of three cancer region detection models were used to generate predicted labels from training and extension sets of $D_{ccRCC}$, $D_{pRCC}$, $D_{chRCC}$. The test set of subtyping consists of two parts: $Test_{patch}$ containing 60832 labeled patches from 65 WSIs for the patch-level results (4-class) evaluation and $Test_{wsi}$ which is the original data (463 WSIs) from TCGA with diagnostic labels for the slide-level (3-class) evaluation. The details of these datasets are shown in Table~\ref{tab1}.

Additionally, four trained annotators were invited to make a complete region annotation on 256 WSIs (123, 88, 45 WSIs for three subtypes). The annotations are cross-reviewed, and then used to generate the region-based annotated data. This data was used to train the fully-supervised model and evaluate the performance of each model.

\subsection{Implementation and Results}
The Resnet-34 architecture \cite{resnet} and the backbone pre-trained on image-net were used for all models in the experiment.

\subsubsection{Cancer Region Detection}

\begin{figure}[t]
\centering
\subfigure[ccRCC]{
\begin{minipage}[b]{0.3\textwidth}
\centering
\includegraphics[width=0.45\textwidth]{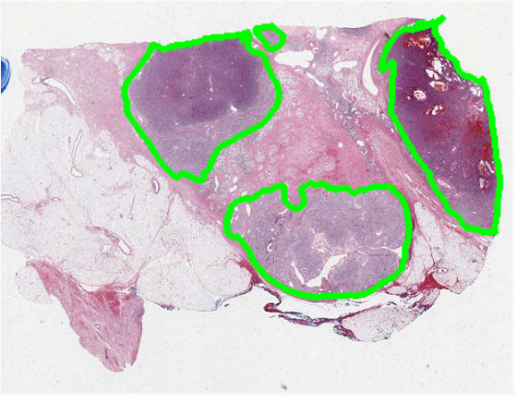}
\includegraphics[width=0.45\textwidth]{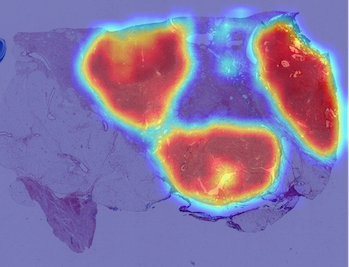}
\end{minipage}}
\subfigure[pRCC]{
\begin{minipage}[b]{0.3\textwidth}
\centering
\includegraphics[width=0.45\textwidth]{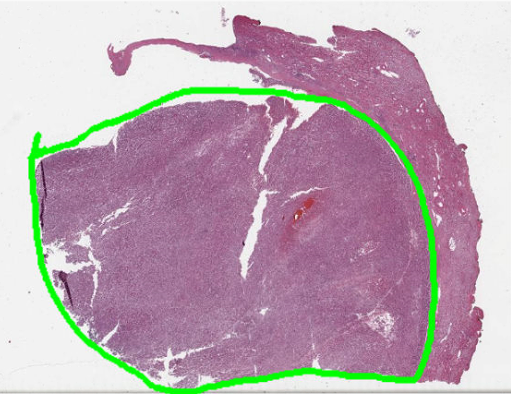}
\includegraphics[width=0.45\textwidth]{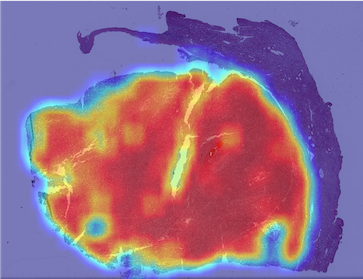}
\end{minipage}}
\subfigure[chRCC]{
\begin{minipage}[b]{0.3\textwidth}
\centering
\includegraphics[width=0.45\textwidth]{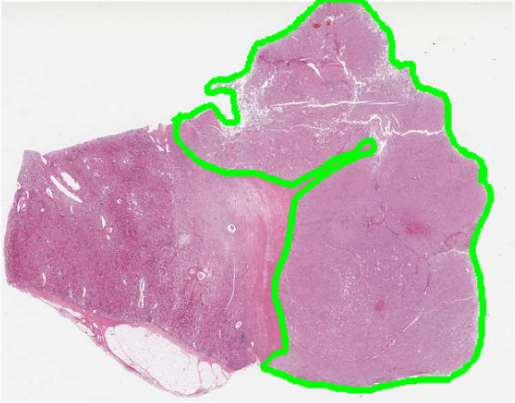}
\includegraphics[width=0.45\textwidth]{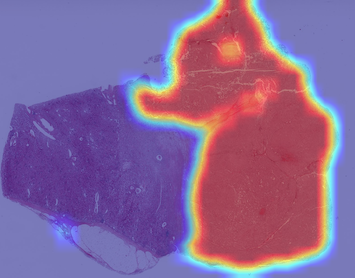}
\end{minipage}}
\caption{Predicted hit-maps of WSIs for three subtypes of RCC. Region-based annotation on the left and predicted hit-maps on the right} \label{fig4}
\end{figure}

\begin{figure}[t]
\centering
\subfigure[AUCs on $D_{ccRCC}$]{
\centering
\includegraphics[width=0.3\textwidth]{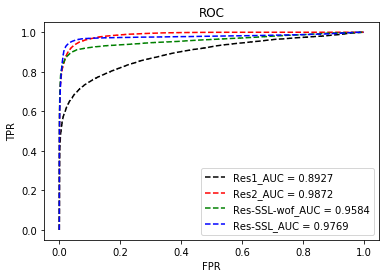}}
\subfigure[AUCs on $D_{pRCC}$]{
\centering
\includegraphics[width=0.3\textwidth]{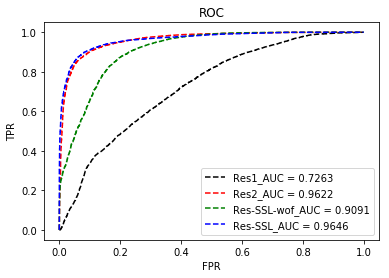}}
\subfigure[AUCs on $D_{chRCC}$]{
\centering
\includegraphics[width=0.3\textwidth]{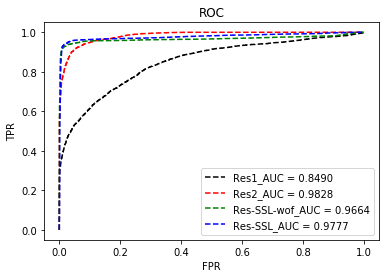}}
\caption{AUC results of $Model_{Res1}$, $Model_{Res2}$, $Model_{Res-SSL-wof}$ and $Model_{Res-SSL}$ on three datasets, $D_{ccRCC}$, $D_{pRCC}$ and $D_{chRCC}$.} \label{fig5}
\end{figure}

Three baseline models ($Model_{Res1}$, $Model_{Res2}$, $Model_{Res-SSL-wof}$) tested on $D_{ccRCC}$, $D_{pRCC}$, $D_{chRCC}$ were used for performance comparison against the proposed framework ($Model_{Res-SSL}$). $Model_{Res1}$ is a Resnet-34 trained on labeled data of training set only. $Model_{Res2}$ is a Resnet-34 trained on segmentation annotated training and extension set, this model is a fully supervised model. $Model_{Res-SSL-wof}$ is a Resnet-34 trained by SSL strategy on training set without fine-tuning. We trained all models with the same optimizer, Adam, with an initial learning rate of 0.001. The learning rate was decreased automatically by the \textit{ReduceLROnPlateau} function with a factor of 10 in Pytorch. All models were trained until convergence, $Model_{Res1}$, $Model_{Res2}$ took 50 epochs, $Model_{Res-SSL-wof}$, $Model_{Res-SSL}$ spent 200-300 epochs, and for $Model_{Res-SSL}$, it cost 5 extra epochs for fine-tuning. We used the cross-entropy loss for training $Model_{Res1}$ and $Model_{Res2}$.

On three datasets, the performance of models shows similar patterns, as illustrated in Fig.~\ref{fig5}. $Model_{Res1}$ shows the worst performance since only a few annotated data is available to the model. $Model_{Res-SSL}$ outperforms $Model_{Res-SSL-wof}$ by about 0.15 AUC on $D_{ccRCC}$ and $D_{chRCC}$, 0.55 AUC on $D_{pRCC}$, where the improvement of performance is due to the fine-tuning. Also, we can see from the results that AUC values of $Model_{Res-SSL}$ and $Model_{Res2}$ are comparable, and gaps between the two models are within $\pm0.1$ AUC. Fig.~\ref{fig4} shows predicted hit-maps of three cases from different datasets by $Model_{Res-SSL}$.

\subsubsection{Subtyping}

\begin{table}[t]
\centering
\caption{Subtyping results}
\label{tab2}
\begin{tabular}{|c|c|c|c|c|}
\hline
\multirow{2}{*}{Model} & \multicolumn{3}{c|}{WSI-wise}                 & Patch-wise    \\ \cline{2-5} 
                       & Precision     & Recall        & F1-Score      & F1-Score      \\ \hline
Res-CE-3class          & 0.78          & 0.82          & 0.79          & -             \\ \hline
Res-CE-4class          & 0.86          & 0.89          & 0.87          & \textbf{0.88} \\ \hline
Res-HB-4class          & \textbf{0.89} & \textbf{0.92} & \textbf{0.91} & 0.87          \\ \hline
\end{tabular}
\end{table}

\begin{table}[t]
\centering
\caption{The confusion matrix of $Model_{CE-4class}$ / $Model_{HB-4class}$ on $Test_{patch}$. Bold numbers indicate reduction of misclassified patches between the two models}
\label{tab3}
\begin{tabular}{|c|c|c|c|c|}
\hline
       & normal        & ccRCC         & pRCC        & chRCC       \\ \hline
normal & 24279 / 24639 & 427 / 259     & 209 / 54     & 55 / 18     \\ \hline
ccRCC  & 1408 / 2342   & 16445 / 16269 & 1177 / \textbf{471}  & 95 / \textbf{43}     \\ \hline
pRCC   & 2116 / 3454   & 154 / \textbf{23}  & 7759 / 6693 & 179 / \textbf{38}    \\ \hline
chRCC  & 332 / 612     & 89 / \textbf{10}   & 516 / \textbf{81}    & 5592 / 5826 \\ \hline
\end{tabular}
\end{table}

For subtyping task, three models were tested. $Model_{CE-4class}$ and $Model_{HB-4class}$ are two Resnet-34 models trained by generated labels with cross-entropy and hybrid loss, respectively. $Model_{CE-3class}$ is a Resnet-34 model trained by diagnostic labels only. All models have the same training process with the models in the previous section.

As shown in Table.~\ref{tab2}, $Model_{CE-4class}$ outperforms $Model_{CE-3class}$ by 8\% on $Test_{wsi}$ in terms of f1-score. We hold ignorance for the normal region accountable for adverse effects on subtyping results. In $Model_{HB-4class}$, the f1-score is 4\% more than $Model_{CE-4class}$ on WSI-wise data, proving the effectiveness of the proposed hybrid loss. Also in Table.~\ref{tab3}, the number of misclassified patches between different subtypes was reduced, but more cancerous patches were classified as normal. Note that predict labels of WSIs are aggregated by patch-level results, except normal class. That is why using hybrid loss causes the model performs a slightly worse on $Test_{patch}$ (Patch-wise in Table.~\ref{tab2}), but better on the final subtyping result.

\section{Conclusion}
Pathologists should not be further burdened by annotation tasks while developing intelligence systems. Streamlined annotation strategies and methods to leverage unlabeled data for digital pathology is urgently needed. In this paper, we introduce a framework with a novel annotation method (Minimal Point-Based annotation) for RCC cancer region detection. It only needs pathologists to select a few points on a limited number of WSIs. This can dramatically reduce annotation efforts over the traditional annotation method. An SSL-based approach for training CNN model can be applied to the annotated data. Our results prove that cancer region detection accuracy of the proposed framework is competitive with a fully supervised learning approach on three major subtypes of RCC. We also propose a subsequent work for RCC subtyping where results show that merely using diagnostic labels and ignoring normal regions for RCC subtype classification is irrational. The proposed framework with hybrid loss has significantly improved accuracy. Our future work will be focusing on developing a clinical-grade diagnostic system for RCC.

\subsubsection{Acknowledgement.}
National Natural Science Foundation of China (61772409); This work has been supported by the National Key Research and Development Program of China (2018YFC0910404); The consulting research project of the Chinese Academy of Engineering (The Online and Offline Mixed Educational Service System for “The Belt and Road” Training in MOOC China); Project of China Knowledge Centre for Engineering Science and Technology; The innovation team from the Ministry of Education (IRT\_17R86); and the Innovative Research Group of the National Natural Science Foundation of China (61721002). The results shown here are in whole or part based upon data generated by the TCGA Research Network: https://www.cancer.gov/tcga.

%
%
%
%

\bibliographystyle{splncs04}
\bibliography{refer}
\end{document}